\documentstyle[preprint,aps]{revtex}
\draft
\tighten
\renewcommand{\vec}[1]{\mbox{\boldmath$#1$\unboldmath}}
\begin{document}

\title{Coulomb dissociation of $^{19}$C}

\author{P. Banerjee}
\address{Theory Group, Saha Institute of Nuclear Physics,
\\1/AF Bidhan Nagar, Calcutta - 700 064, INDIA}
\date{\today}
\maketitle

\begin{abstract}
We aim to re-investigate the structure of the neutron rich exotic nucleus $^{19}$C
through studies of its Coulomb breakup reactions, assuming that its excitation 
is to states in the low energy continuum. The method used 
retains all finite-range effects associated with the interactions between the 
breakup fragments and can use realistic wave functions for this possibly halo 
nucleus. We apply the method to compute the longitudinal momentum distribution 
of $^{18}$C, exclusive neutron angular distribution and relative energy spectrum of 
$^{19}$C following Coulomb breakup of $^{19}$C on heavy targets at beam energies 
below 100 MeV/nucleon. The calculated results are compared with recently 
available experimental data. In the majority of the few cases, the data favour 
$^{18}$C(0$^+)\otimes$1$s_{1/2}$ as the probable ground state configuration of
$^{19}$C, with a one-neutron separation energy of 0.53 MeV. However, it appears
that the existing experimental data do not allow to draw a clear conclusion 
about the ground state configuration of this nucleus. 
\end{abstract}


\section{Introduction}

The neutron rich nucleus $^{19}$C has drawn much attention recently as candidate
for having a one-neutron halo structure \cite{1,2}. This can be easily anticipated
due to its very small neutron binding energy (of the order of a few hundred keV).
However, halo formation not only depends upon the small separation energy of the
valence neutron. The spin-parity and the configuration of the single particle
state of the valence neutron also play important roles in the formation of the halo. 
The case in which
$^{19}$C has a $J^\pi$=1/2$^+$ ground state (with a dominant $^{18}$C(0$^+)\otimes$1$s_{1/2}$ 
single particle configuration \cite{3}) favours the halo formation. On the other
hand, the case of a $J^\pi$=5/2$^+$, expected from standard shell model ordering, 
prevents the halo from being formed due to the large centrifugal barrier associated
with a 0$d_{5/2}$ orbit. The $J^\pi$=5/2$^+$ may also result when the dominant 
single particle configuration in the ground state is $^{18}$C(2$^+)\otimes$1$s_{1/2}$. 
In this case, the binding energy of the valence neutron is effectively increased
by the excitation energy of 1.62 MeV of the excited (2$^+$) state of the $^{18}$C
core, thereby also hindering the halo formation. A $J^\pi$=3/2$^+$ is also possible
with this configuration for the ground state, as considered in recent model 
calculations \cite{4}.

In order to probe the structure of $^{19}$C, only a few experiments on the 
dissociation of $^{19}$C have been carried
out so far. Some of them are inclusive measurements, in which observables
related to either the $^{18}$C core or the valence neutron have been measured 
following the breakup of $^{19}$C. The longitudinal momentum distribution of 
$^{18}$C has been measured at incident energy of 88 MeV/nucleon at MSU on Be and 
Ta targets \cite{1}, where a very narrow momentum width of around 42 MeV/c FWHM 
(full width at half maximum) has been reported. A similar
experiment, conducted at GSI at very high incident energy of around 910 MeV/nucleon
on a carbon target, shows a rather broad width of 69 $\pm$ 3 MeV/c \cite{5}. The 
neutron momentum distribution, measured in a core breakup reaction of $^{19}$C at 30 MeV/nucleon 
energy on Ta target at GANIL, shows a narrow component of 42 $\pm$ 17 MeV/c FWHM \cite{2}. This
is in contradiction with the observed broad width of 120 $\pm$ 18 MeV/c FWHM of
the neutron angular distribution measured at GANIL in an elastic breakup of 
$^{19}$C \cite{6}. 

Very recently the relative energy spectrum of $^{19}$C has
been measured in its Coulomb dissociation at 67 MeV/nucleon incident energy on Pb target in a kinematically 
complete experiment at RIKEN \cite{7}. The RIKEN data, when compared with
semi-classical calculations \cite{7}, support a dominant 1$s_{1/2}$ 
configuration of $^{19}$C halo wave function. The MSU data, on the other hand,
suggest an $s$-wave neutron around the 2$^+$ excited state of $^{18}$C \cite{1}. 
Recent theoretical
calculations on the Coulomb breakup of $^{19}$C \cite{8} indicate the 
possibility of an $s$-wave neutron coupled to the 2$^+$ excited state of $^{18}$C, 
or a $d$-wave neutron, or a coherent superposition of these two configurations 
as far as the recent GANIL neutron angular distribution data are 
concerned. The total one-neutron removal cross sections in the Coulomb dissociation 
of $^{19}$C on heavy targets 
have been calculated in \cite{8}. These cross sections, when compared with
experimental data on one-neutron removal cross sections measured at GANIL \cite{6}
and MSU \cite{1}, show that the configuration of a 1$s_{1/2}$ neutron coupled
to the 0$^+$ $^{18}$C core has non-significant contribution to the ground state 
structure of $^{19}$C \cite{8}. However, very recent calculations of reaction cross 
sections of $^{19}$C on a carbon target at relativistic energies \cite{23}, when
compared with experimental data \cite{24}, show the dominance of the $^{18}$C(0$^+)
\otimes$1$s_{1/2}$ ground state configuration. In view of all these discrepancies, 
it has not been possible to draw any definite conclusion on the structure of 
$^{19}$C. 

The difficulties in drawing conclusion about the structure of $^{19}$C might arise 
also due to lack of precise knowledge of the 
one-neutron separation energy in $^{19}$C. This quantity is very crucial as far as the
property of the neutron halo is concerned. The value used in
most of the calculations to date is 240 keV. This is the average over previous
four measurements of this quantity \cite{9,10,11,12}. The recent measurement
of the relative energy spectrum of $^{19}$C in its breakup on Pb at RIKEN has 
led to an indirect determination of the
$^{19}$C one-neutron separation energy \cite{7}, which is different from that
mentioned above (see below).

In this paper, we attempt to re-investigate the structure of $^{19}$C
through studies of breakup reactions induced by it in the Coulomb field of a heavy target 
at beam energies below 100 MeV/nucleon. We compare our results with all the sparsely
available differential breakup data and try to throw light on its structure. We follow the 
theoretical formalism which was first described in \cite{13} for the Coulomb 
breakup of a light weakly bound two-body composite nucleus $a$ consisting of a 
charged core $c$ and a 
neutral valence particle $v$ on target $t$ at energies of a few tens of MeV per
nucleon and above. 
There are two approximations used in this theory - that the dominant projectile
breakup configurations excited are in the low-energy continuum and that the
valence particle does not interact with the target. The theory is fully
quantum mechanical and is also non-perturbative. The method retains all 
finite-range effects associated with the interactions among the breakup 
fragments and includes the initial and final state interactions to all orders. 
It allows the use of realistic wave functions to describe the halo nuclei.

We describe the theoretical formalism in section 2. Structure models are
discussed in section 3. Section 4 deals with results and discussions. Finally, 
we conclude in section 5.

\section{Formalism}

The transition amplitude for the elastic Coulomb breakup reaction $a + t \rightarrow
c + v + t$, in the c.m. frame, is given by (Fig.1)
\begin{equation}
T_{\sigma _c\sigma _v;\sigma _a}=\langle \chi^{(-)} (\vec{k}_c, \vec{R}_c)
{\cal S}_{\sigma _c} e^{i\vec{k}_v\cdot\vec{R} _v}{\cal S}_{\sigma _v}\vert
V_{cv}\vert \Psi ^{(+)}_{\vec{k}_a{\sigma _a}} (\vec{r},\vec{R})\rangle~,
\label{tmat}
\end{equation}
where ${\cal S}_{\sigma _c}$ and ${\cal S}_{\sigma _v}$ are the core and
valence particle internal wavefunctions with $\sigma_c$ and $\sigma_v$ their
spin projections. $\hbar\vec{k}_c$ and $\hbar\vec{k}_v$ are the asymptotic
momenta of these fragments, conjugate to $\vec{R}_c$ and $\vec{R}_v$,
respectively, and $\chi^{(-) }$ is an in-going waves Coulomb distorted wave
function describing the $c$--$t$ relative motion in the final state. Since it
is assumed that $V_{vt}=0$ the valence particle is described by a plane wave in
the final state.

Following the adiabatic approximation of ref.\ \cite{14}, the exact three-body
scattering wave function $\Psi ^{(+)}_{\vec{k}_a{\sigma _a}} (\vec{r},\vec{R})$
separates in the variables $\vec{R}_c$ and $\vec{r}$, namely
\begin{equation}
\Psi ^{(+)}_{\vec{k}_a\sigma _a}
(\vec{r},\vec{R}) \approx \Psi_{\vec{k}_a{\sigma _a}}^{\rm AD}(\vec{r},\vec{R})
= \Phi _{a\sigma _a}(\vec{r}) e^{i\gamma\vec{k}_a\cdot \vec{r}} \chi
^{(+)}(\vec{k}_a,\vec{R}_c)~.
\label{adwf}
\end{equation}
Here $\chi ^{(+)}$ is a Coulomb distorted wave representing projectile's
motion in the incident channel, evaluated at the core coordinate $\vec{R}_c$
and $\gamma = m_v/(m_c + m_v)$.  

The projectile ground state wave function is given by 
\begin{eqnarray}
\Phi _{a\sigma _a}(\vec{r})=\sum_{l\mu jm\sigma_c'\sigma_v'} \langle s_c \sigma
_c'jm \vert s_a\sigma _a\rangle\langle l\mu s_v\sigma _v'\vert jm \rangle
\Phi_{a}^{l\mu}(\vec{r}){\cal S}_{\sigma _c'}{\cal S}_{\sigma _v'}~,
\end{eqnarray}
where $\Phi _{a}^{l\mu}(\vec{r})= i^l u_l(r) Y_{l\mu}(\hat{\vec{r}})$, the
$u_l$ are radial wavefunctions, and the $Y_{l\mu}$ are the spherical
harmonics.  Since the only distorting interaction $V_{ct}$ is assumed to be central,
the integrations over spin variables can be carried out in Eq.\ (\ref{tmat}).
The required approximate transition amplitude can then be expressed as
\begin{eqnarray}
T^{\rm AD}_{\sigma _c\sigma _v;\sigma _a}=\sum_{l\mu j m} \langle s_c\sigma _cj
m \vert s_a\sigma _a\rangle\langle l\mu s_v\sigma _v\vert jm \rangle 
\beta^{AD}_{l\mu}~,
\end{eqnarray}
where the reduced transition amplitude $\beta^{AD}_{l\mu}$ is 
\begin{eqnarray}
\beta^{\rm AD}_{l\mu}=\langle\chi ^{(-)}(\vec{k}_c,
\vec{R}_c) e^{i\vec{k}_v \cdot \vec{R}_v} \vert V_{cv}\vert \Phi
_{a}^{l\mu}(\vec{r}) e^{i\gamma \vec{k}_a \cdot \vec{r}} \chi
^{(+)}(\vec{k}_a,\vec{R}_c) \rangle~.
\end{eqnarray}
Since $\vec{R}_v= \alpha\vec{R}_c + \vec{r}$ (Fig.1), where $\alpha = m_t/(m_t
+m_c)$, then without further approximation the entire adiabatic amplitude now
separates exactly in the coordinates $\vec{R}_c$ and $\vec{r}$, as
\begin{eqnarray}
\beta^{\rm AD}_{l\mu}&=&\langle e^{i\vec{q}_v\cdot \vec{r}}\vert V_{cv}\vert
\Phi _{a}^{l\mu}(\vec{r}) \rangle \langle \chi ^{(-)}(\vec{k}_c,
\vec{R}_c)e^{i\alpha\vec{k}_v\cdot \vec{R}_c} \vert \chi
^{(+)}(\vec{k}_a,\vec{R}_c) \rangle~\nonumber \\ 
&=&\langle \vec{q}_v\vert V_{cv}\vert \Phi _{a}^{l\mu}\rangle
\langle \chi ^{(-)}(\vec{k}_c);\alpha\vec{k}_v\vert \chi
^{(+)}(\vec{k}_a) \rangle~.\label{betad}
\end{eqnarray}
The momentum $\vec{q}_v$ appearing in the first term is $\vec{q}_v
=\vec{k}_v-\gamma\vec{k}_a$.

Here the structure information about the projectile
is contained only in the first term, the vertex function, denoted by 
$D(\vec{q}_v) = D_l(q_v)Y_{l\mu}(\hat{\vec{q}}_v)$, where
\begin{eqnarray}
D_l(q)= 4\pi\int^{\infty}_0 dr r^2 j_l(qr)V_{cv}(r) u_l(r)~.
\end{eqnarray}
The second factor is associated with the dynamics of the reaction only, which
is expressable in terms of the bremsstrahlung integral \cite{15}.

The triple differential cross section for the elastic breakup reaction is
\begin{eqnarray}
{d^3\sigma \over dE_c d\Omega _cd\Omega _v}={2\pi\over \hbar v_a}\left\{{1\over
2s_a + 1}\sum_{\sigma _c\sigma _v\sigma _a}\vert T^{\rm AD}_{\sigma _c\sigma _v;
\sigma _a}\vert ^2\right\} \rho (E_c,\Omega _c,\Omega _v)  ~, \label{trip}
\end{eqnarray}
or, upon carrying out the spin projection summations,  
\begin{eqnarray}
{d^3\sigma \over dE_c d\Omega _cd\Omega _v}={2\pi\over \hbar v_a}\left\{\sum_{l
\mu}\frac{1}{(2l + 1)}\vert \beta^{\rm AD}_{l\mu}\vert^2 \right\} \rho(E_c,
\Omega _c,\Omega _v)~.
\end{eqnarray}
Here $v_a$ is the $a$--$t$ relative velocity in the entrance channel.
The phase space factor $\rho (E_c,\Omega _c,\Omega _v)$ appropriate to the
three-body final state is \cite{16,17}
\begin{eqnarray}
\rho (E_c,\Omega _c,\Omega _v) = {h^{-6}m_tm_cm_vp_cp_v\over m_v + m_t
-m_v\vec{p}_v\cdot(\vec{P} -\vec{p}_c)/p_v^2} \label{phas}
\end{eqnarray}
where, for the differential cross section in the laboratory frame, $\vec{P}$,
$\vec{p}_c$ and $\vec{p}_v$ are the total, core, and valence particle momenta in
the laboratory system.

The neutron angular distribution is
obtained by integrating the above triple differential cross section with
respect to solid angle and energy of the core fragment. 
The core three dimensional momentum distribution $d^3\sigma \over dp_{x,c} 
dp_{y,c} dp_{z,c}$ is related to its energy distribution cross section by
\begin{eqnarray}
{d^3\sigma \over dp_{x,c} dp_{y,c} dp_{z,c}}
 = {1\over m_c\sqrt {2m_cE_c}}{d^2\sigma \over dE_c d\Omega _c}
\end{eqnarray}
${d^2\sigma \over dE_c d\Omega _c}$ can be readily obtained from the triple
differential cross section above (Eq. (8)) by integration with respect to the solid
angle of the valence particle. Starting from the three dimensional momentum 
distribution, the parallel momentum distribution of the heavy charged fragment 
$c$ can be obtained by integration over the transverse momentum components. 
To calculate the $c$--$v$ relative energy spectrum 
$d\sigma/dE_{cv}$ from the above triple differential cross section, 
we follow ref.\cite{8}. Since the excitation
energy $E_{ex}$ is equal to $E_{cv} + \epsilon _0$, where $\epsilon _0$ is the 
one-neutron separation energy, the relative energy spectrum is also equal to
the excitation energy spectrum $d\sigma/dE_{ex}$.

\section{Structure Models}

We have considered 
the following configurations for the valence neutron in $^{19}$C: (a) a 1$s_
{1/2}$ state bound to a 0$^+$ $^{18}$C core by 0.24 MeV, (b) a 1$s_{1/2}$
state bound to a 2$^+$ $^{18}$C core by 1.86 MeV and (c) a 0$d_{5/2}$ 
state bound to a 0$^+$ $^{18}$C core by 0.240 MeV. Recently, another option 
for the ground state structure of $^{19}$C has been proposed \cite{7}. We 
consider it as option (d). In this case, the ground state is a moderately 
dominant 1$s_{1/2}$ 
neutron configuration (with a spectroscopic factor $S$ = 0.67) which is bound to a
0$^+$ core with 0.53 MeV. The binding potentials in all cases are taken to be of 
Woods-Saxon type with the radius and diffuseness parameters as 1.15 fm and
0.5 fm respectively. Their depths have been calculated to reproduce the 
respective binding energies. The rms sizes with the above options are 3.45 fm, 
3.00 fm, 2.96 fm and 3.20 fm respectively. The rms size used for the $^{18}$C 
core is 2.9 fm \cite{18}. These different wave functions of $^{19}$C give rise
to different vertex functions (Eq. (7)) and consequently, different Coulomb 
breakup cross sections \cite{8}.

\section{Results and discussions}

In the very recent kinematically complete measurement of the Coulomb 
dissociation of $^{19}$C on Pb at 67 MeV/A at RIKEN \cite{7}, the breakup
fragments have been detected within a narrow forward cone of opening angle
of 2.5$^\circ$. The grazing angle is 2.6$^\circ$. The excitation (relative) energy spectrum
has been constructed and the dipole strength distribution deduced therefrom. The
excitation energy spectrum and the extracted dipole strength distribution
have strong peaks at excitation energy of almost 800 keV, then it
dumps to almost zero at $E_{ex}$ = 2 MeV. The peak height of the excitation 
(relative) energy spectrum is $\simeq$1 barn/MeV. A neutron separation energy of 
530 keV has been derived indirectly and independently by the analysis of the 
angular distribution of the $^{19}$C centre of mass system. 
The shape of the relative energy spectrum ($d\sigma \over dE_{rel}$ as a function 
of $E_{rel}$), as calculated by
these authors, has been found in good agreement with that resulting from a 
ground state spin-parity
assignment of 1/2$^+$, corresponding to a 1$s_{1/2}\otimes ^{18}$C(0$^+$)
configuration. The spectral amplitude has been used to extract the
ground state spectroscopic factor. This has
come out to be 0.67. The total Coulomb breakup cross 
section has been estimated to be 1.2$\pm$0.11 b. We have 
calculated the relative energy spectra of $^{19}$C on Pb up to $E_{rel}$ = 4
MeV at the same beam 
energy using the above four configurations (Fig.2). The angular integration
for the centre of mass of the $^{18}$C + $n$ system has been done up to 2.5$^{\circ}$.
The spectrum in the case
(d) only has peak position agreeing with that of the 
experimental data, the peak height being slightly smaller than that of the
experimental data. The relative energy spectra, calculated with other three
wave functions, have got significantly different magnitudes and different peak
positions (Fig.2). But the shape of
the spectrum in the case (d) is somewhat different from that of the experimental
one for $E_{rel}>$0.4 MeV. The calculated spectrum is smaller than the experimental
one in this range. The total Coulomb breakup cross section of 0.75 b, computed
in this case, is also less than the value quoted above. It is to be noted 
that we have not used the spectroscopic factor of 0.67 to get this cross 
section (and the relative energy spectrum). 
Rather we have used $S$ = 1. It is evident that use of $S$ = 0.67 results in
even smaller relative energy spectrum and total Coulomb part of one-neutron
removal cross section. It is worth mentioning that the sensitivity of the extracted
spectroscopic factor of 0.67 to the geometry of the assumed (Woods-Saxon) $^{18
}$C + $n$ binding potential was not clarified in the calculations in \cite{7}. 
The parameters of the potential were also not specified in \cite{7}. Also, the
relative energy spectrum in pure Coulomb dissociation was obtained by subtracting
the nuclear breakup contribution, calculated approximately by the authors of ref.\cite{7}, 
from the experimental relative energy spectrum. An accurate estimate of this 
nuclear breakup contribution as well as Coulomb-nuclear interference is necessary
for arriving at any definite conclusion from comparison with the extracted 
Coulomb breakup contribution. We infer that the configuration proposed for 
$^{19}$C in \cite{7} is supported by our calculations only to some extent. 

The longitudinal or parallel momentum distribution (PMD) of the heavy charged fragments in the
breakup of the neutron halo nuclei is known to give information on the halo 
structure \cite{1,8,19,20,21,22}. The transverse momentum distributions, on 
the other hand, are substantially affected by the reaction mechanism 
\cite{8,21,22}. For breakup of 
nuclei with very small separation energies of the valence neutron(s)
and on high $Z$ targets, the PMDs are expected to have a large contribution from 
Coulomb breakup \cite{1,8}. We have calculated the parallel momentum distribution 
$d\sigma \over dp_z$ as a function $p_z$ of $^{18}$C following 
breakup of $^{19}$C on a Ta target at 88 MeV/nucleon beam energy. This has been
recently measured at MSU \cite{1}. Calculations with the first three options on
the $^{19}$C wave function have been reported earlier \cite{25}. We use these
results here and present them along with the result of calculation with option (d).
The widths and absolute magnitudes of the
PMDs are quite different in the four cases (Fig.3). The widths (FWHM) 
are 27, 71, 83 and 41 MeV/c respectively for the options (a), (b), (c) and (d)
cited above for the $^{19}$C ground state wave function. The MSU data do not 
have the absolute magnitudes of the momentum distributions. The published width
of 41$\pm$3 MeV/c, deduced from these data, favours option (d), although 
these data have limited statistics (Fig.4). This is 
in contradiction to the observation of Ridikas {\em et al.} \cite{4}, who
investigated $^{19}$C in a core-plus-neutron coupling model with a deformed
Woods-Saxon potential for the neutron-core interaction. They obtained a
reasonable fit to the $^{18}$C PMD when $J^{\pi}$ = 3/2$^+$, 5/2$^+$ was
assumed, and when the wave function had an appreciable amount of $s$-motion
coupled to the 2$^+$ state of $^{18}$C. But this calculation used a one-neutron
separation energy of 0.24 MeV. With a different choice of separation energy
of 0.5 MeV and with $^{18}$C(0$^+)\otimes$1$s_{1/2}$ ground state configuration,
they were able to reproduce such a narrow width. This one-neutron separation 
energy is close to that used in option (d) of our calculations, the ground state
configuration is the same as that in (d). It is to be noted that the theoretical
calculations in \cite{4} were concerned with the breakup of $^{19}$C on a light
target (Be), for which Coulomb breakup is not the dominant reaction mechanism.

Using the options (a), (b) and (c) above for $^{19}$C ground state wave 
function and following the same theoretical formalism used here, Coulomb breakup 
calculations on the recent measurement \cite{6} of 
the neutron angular distribution following $^{19}$C elastic breakup on 
Ta at 30 
MeV/nucleon were reported earlier \cite{8}. This observable also gives a direct
indication of the ground state structure. We have repeated the calculations 
with the wave function (d) for $^{19}$C. The magnitude of this cross section at
forward angles is expected to have a significant contribution from the Coulomb
breakup mechanism calculated here. Fig.5 shows the calculated angular distributions
($d\sigma \over d\Omega _n$ as a function of $\theta _n$)
with the above four configurations of $^{19}$C. The data in \cite{6}, reproduced
in Fig.5, show a broad neutron distribution with a 
FWHM of 120$\pm$ 18 MeV/c. The cross section magnitude is seen to be
$\approx 1.5$ b/sr at forward angles. Of our Coulomb breakup calculations with
the above four options for the ground state configuration of $^{19}$C, only
the model in which the ground state is described as an $s$-wave neutron coupled
to a core excited state comes close, both in magnitude and shape, to the data.
In this case, we have calculated the cross section for detection of the $^{18}$C core
in the excited (2$^+$) state. A coherent superposition of $^{18}$C$(0^+)
\otimes 0d_{5/2}$ and  $^{18}$C$(2^+) \otimes 1s_{1/2}$ configurations is also
allowed, for a $5/2^+$ ground state, and would lead to an incoherent
superposition of the lower two curves in Fig.5.

\section{Conclusions}

In conclusion, we have studied Coulomb dissociation of the neutron rich exotic
nucleus $^{19}$C and compared our calculations with the existing experimental 
results in order to probe its possible halo structure. The calculations are 
performed within an approximate quantum mechanical theoretical model of elastic
Coulomb breakup, which makes
the following assumptions: (i) only the charged core interacts with the target
via a point Coulomb interaction and (ii) that the important excitations of the
projectile are to the low-energy continuum, and so can be treated adiabatically.
The method permits a fully finite-range treatment of the projectile vertex and
includes initial and final state interactions to all orders. The comparison 
between the calculated
and experimentally measured parallel momentum distributions of the breakup 
fragment $^{18}$C and the relative energy spectra of $^{19}$C gives some support
to a dominant $^{18}$C(0$^+)\otimes$1$s_{1/2}$ ground state configuration with
1/2$^+$ spin-parity and therefore, to a neutron halo structure. These calculations
use a one-neutron separation energy of 0.53 MeV. But this configuration is in 
contradiction with $^{18}$C(0$^+)\otimes$0$d_{5/2}$ or $^{18}$C(2$^+)\otimes$1$s_{1/2}$ or a 
superposition of these two ground state configurations (with $J^{\pi}$ = 5/2$^+$) necessary
to account for the neutron angular distribution (at forward angles) measured at
GANIL, which results from the elastic breakup of $^{19}$C. The one-neutron binding
energy of 1.86 MeV/0.24 MeV is also different in this case. We see that in the 
majority of the available few cases, the data favour $^{18}$C(0$^+)\otimes$1$s_
{1/2}$ as the probable ground state configuration of $^{19}$C, with a one-neutron 
separation energy of 0.53 MeV. In view of the insufficiently available data, we 
feel that more good quality experimental measurements will prove useful in 
arriving at a more definite conclusion.

\acknowledgments 
The author would like to thank Prof. R. Shyam for making valuable comments.

\begin{figure}
\caption{Coordinate system adopted for the core, valence particle and
target three-body system.}
\end{figure}

\begin{figure}
\caption{Calculated relative energy spectra in the Coulomb dissociation of 
$^{19}$C on a Pb target at 67 MeV/nucleon. The dotted, short dashed and long
dashed curves are results of multiplication by 0.16, 10 and 10 respectively of the
actual calculations. The experimental data have been taken from \protect\cite{7}.}
\end{figure}

\begin{figure}
\caption{Calculated parallel momentum distributions of $^{18}$C from Coulomb breakup of
$^{19}$C on Ta at 88 MeV/nucleon. The short dashed and long dashed curves result
when actual calculations in these two cases have been multiplied by a factor of
55. We have not used $S$ = 0.67 in case of the solid curve, as the experimental
data (not shown here) do not have absolute magnitudes.}
\end{figure}

\begin{figure}
\caption{Calculated parallel momentum distribution of $^{18}$C from Coulomb breakup of
$^{19}$C on Ta at 88 MeV/nucleon with option (d) (see text). The peak of the 
calculation has been normalized to that of the data (in arbitrary units),
taken from \protect\cite{1}. The centroid of the data has been shifted to 
compare the widths.}
\end{figure}

\begin{figure}
\caption{Calculated neutron angular distributions from Coulomb breakup of
$^{19}$C on Ta at 30 MeV/nucleon. The experimental data have been taken from 
\protect\cite{6}. The solid curve has been obtained with $S$ = 0.67.}
\end{figure}
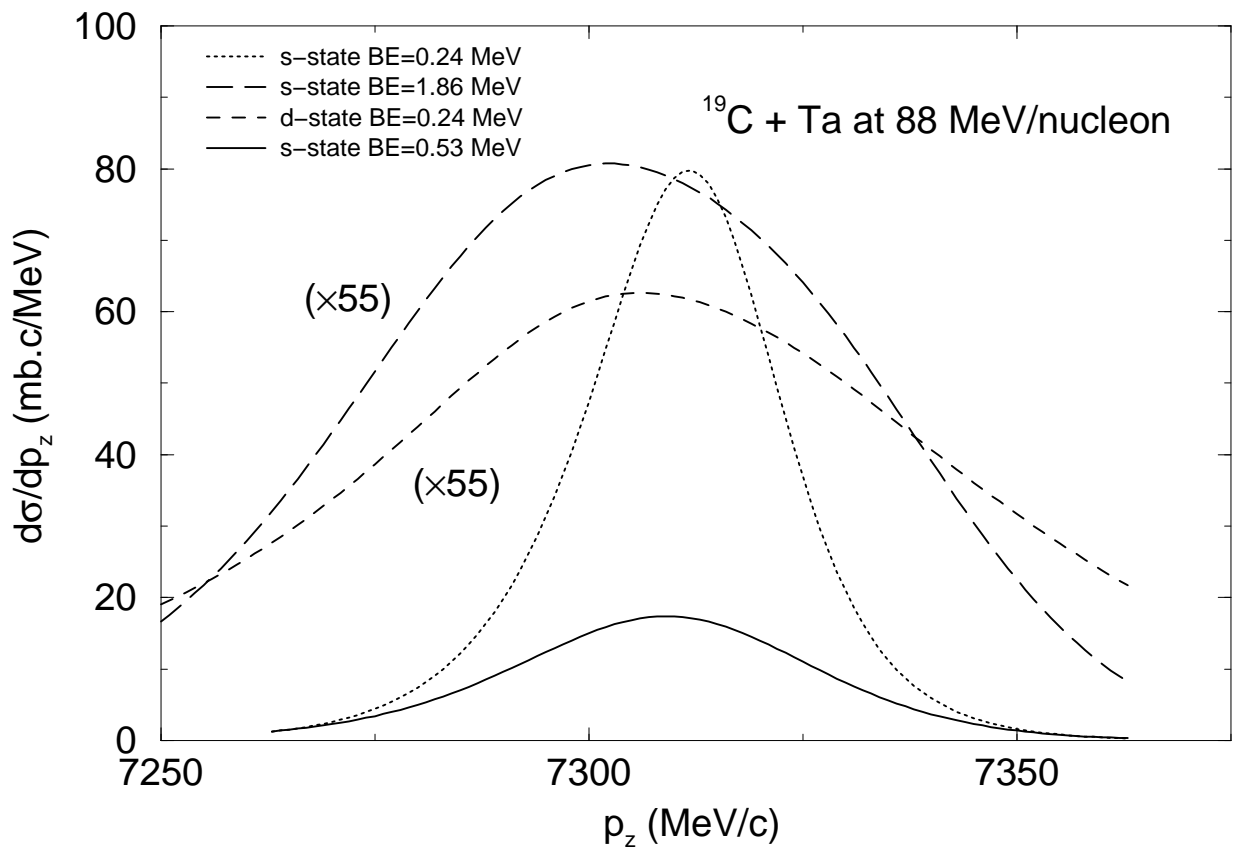
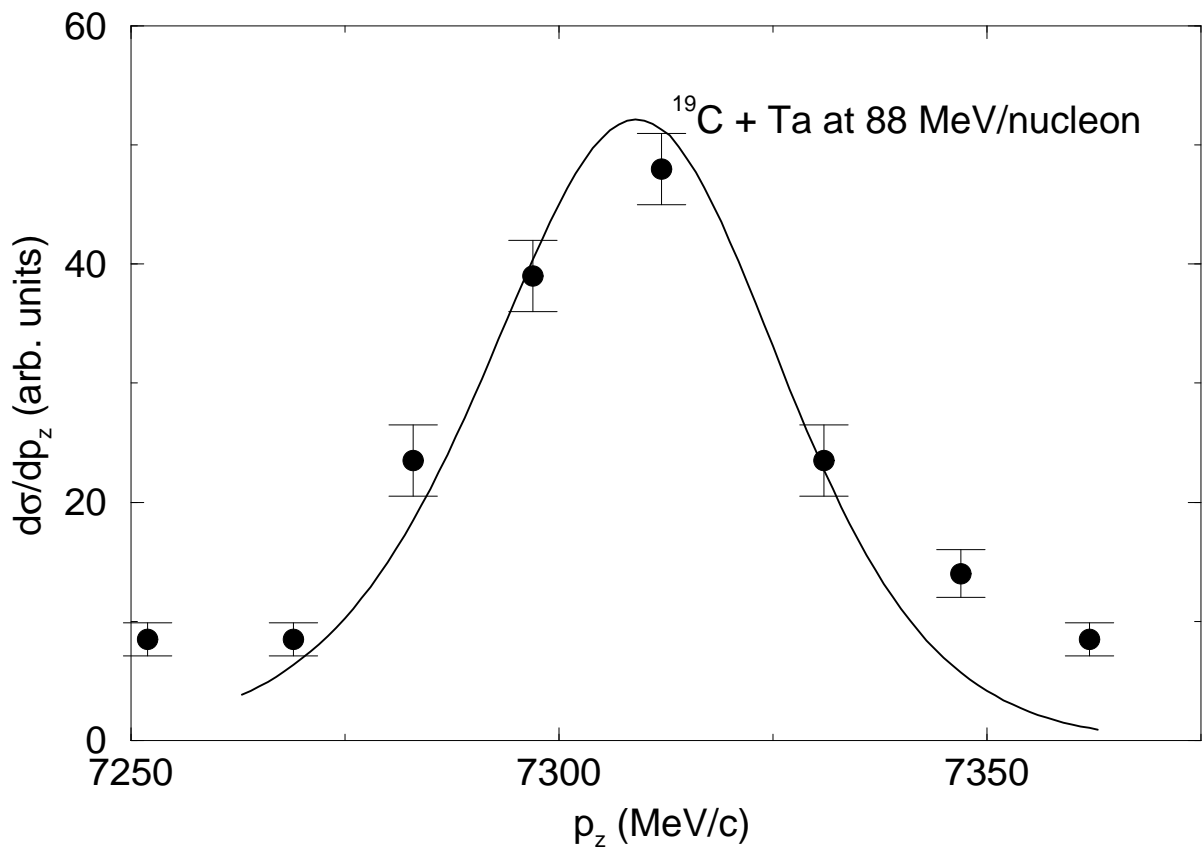
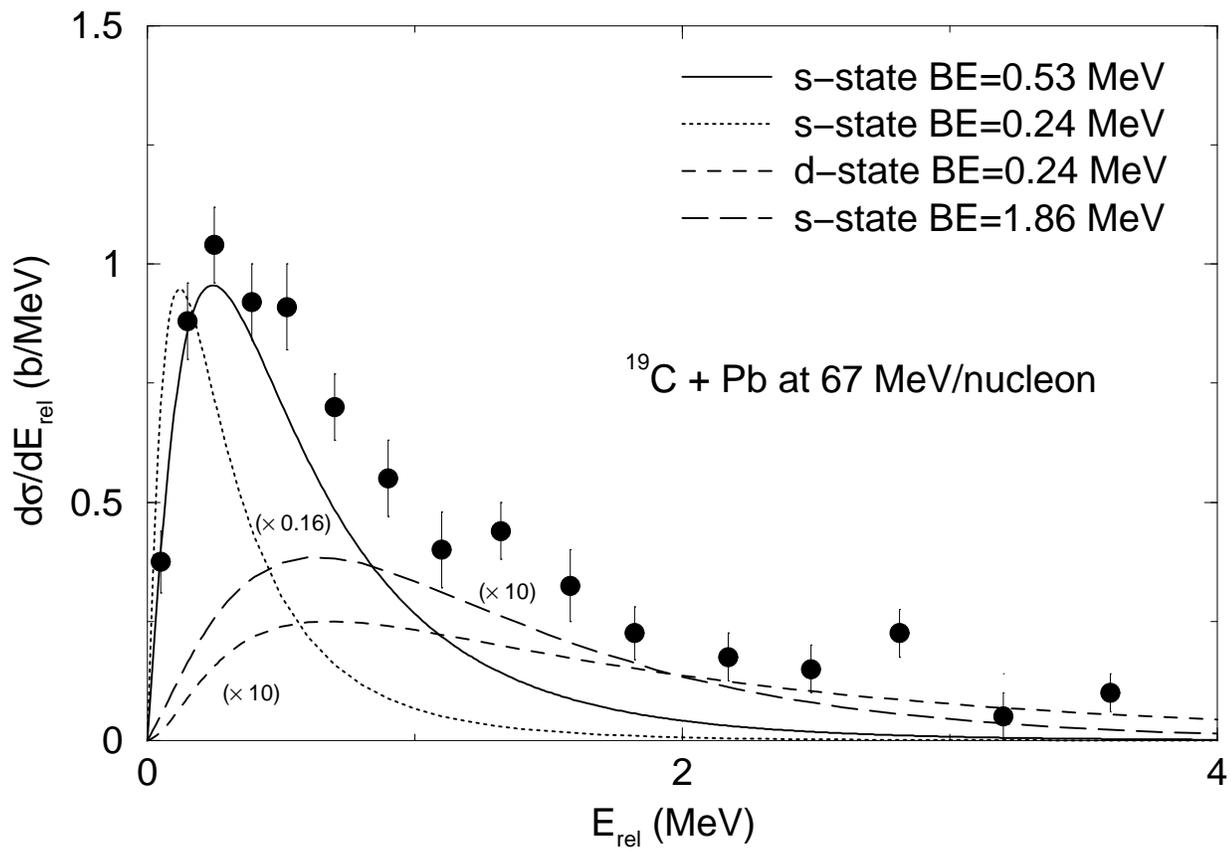
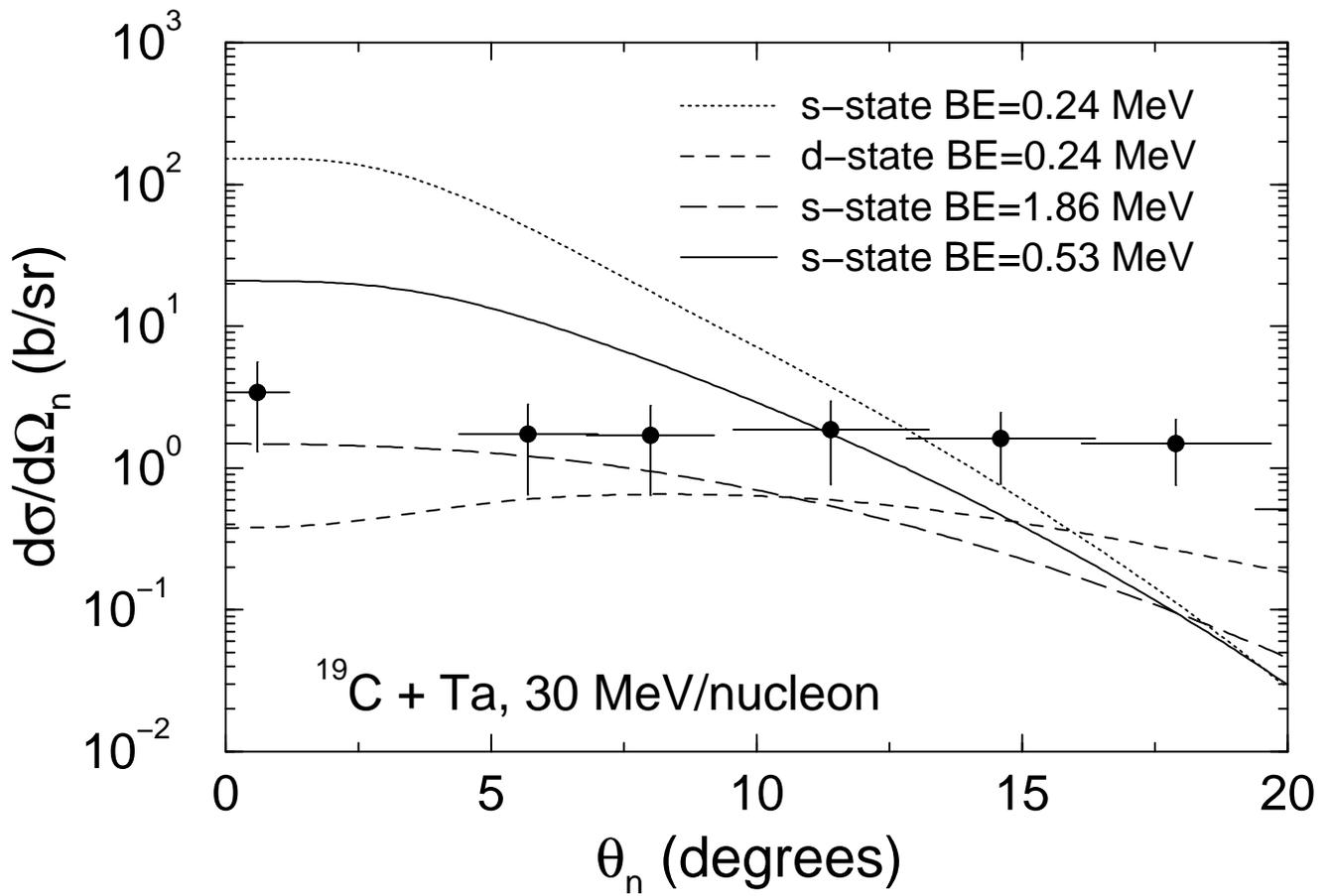
\end{document}